# Lung Cancer Diagnosis Using Deep Attention Based Multiple Instance Learning and Radiomics

**Running Title:** Lung Cancer Diagnosis with MIL

Junhua Chen [a,1], MS; Haiyan Zeng [1], MD; Chong Zhang [1], MS; Zhenwei Shi [1], PhD; Andre Dekker [1], PhD; Leonard Wee [1,*], PhD; Inigo Bermejo [1,*], PhD

[1]. Department of Radiation Oncology (MAASTRO), GROW School for Oncology and Developmental Biology, Maastricht University Medical Centre+, Maastricht, 6229 ET, Netherlands

**Address of Corresponding author:** Department of Radiation Oncology (MAASTRO), GROW School for Oncology and Developmental Biology, Maastricht University Medical Centre+, Maastricht, 6229 ET, Netherlands

[a] Corresponding author, Tel.: +31 0684 6149 49 E-mail address: j.chen@maastrichtuniversity.nl

* equally senior author

**Abstract**

**Background:** Early diagnosis of lung cancer is a key intervention for the treatment of lung cancer in which computer aided diagnosis (CAD) can play a crucial role. Most published CAD methods perform lung cancer diagnosis by classifying each lung nodule in isolation. However, this does not reflect clinical practice, where clinicians diagnose a patient based on a set of images of nodules, instead of looking at one nodule at a time. Besides, the low interpretability of the output provided by these methods presents an important barrier for their adoption.

**Method:** In this article, we treat lung cancer diagnosis as a multiple instance learning (MIL) problem, which better reflects the diagnosis process in the clinical setting and provides higher interpretability of the output. We selected radiomics as the source of input features and deep attention-based MIL as the classification algorithm. The attention mechanism provides higher interpretability by estimating the importance of each instance in the set for the final diagnosis. In order to improve the model's performance in a small imbalanced dataset, we propose a new bag simulation method for MIL.

**Results and Conclusion:** The results show that our method can achieve a mean accuracy of 0.807 with a standard error of the mean (SEM) of 0.069, a recall of 0.870 (SEM 0.061), a positive predictive value of 0.928 (SEM 0.078), a negative predictive value of 0.591 (SEM 0.155) and an area under the curve (AUC) of 0.842 (SEM 0.074), outperforming other MIL methods. Additional experiments show that the proposed oversampling strategy significantly improves the model's performance. In addition, our experiments show that our method provides a good indication of the importance of each nodule in determining the diagnosis, which combined with the well-defined radiomic features, make the results more interpretable and acceptable for doctors and patients.

# 1. Introduction

According to the statistics from the World Health Organization (WHO), lung cancer is the most frequently diagnosed malignant carcinoma and the leading cause of cancer death worldwide, accounting for an estimated 2.09 million deaths in 2018 [1][2]. Early diagnosis and treatment can reduce a lung cancer patient's mortality significantly. A plausible method for early lung cancer diagnosis is the routine use of low dose computed tomography (CT) scans [3]. To date, radiologists typically need to visually inspect CT scans slice by slice, which is costly and time-consuming as well as susceptible to human error [4][5]. Computer aided diagnosis (CAD) for rapid early lung nodules classification based on low-dose CT imaging has therefore attracted much attention from researchers during the last decades [6][7].

The development of CAD for lung nodules classification has reached new peaks in last decade mainly due to breakthroughs in deep learning neural networks [8] and its application to a wide range of medical image analysis tasks. Several deep learning-based lung nodule classification methods have been proposed in recent years, with steadily improving state-of-art performance. Shen et al. [9] developed a multi-scale convolutional neural network (CNN) to extract features (referred to as 'deep features' [10] in the literature) then applied a supervised random forest classifier to the deep features, reporting an accuracy of 86%. Xie et al. [11] combined handcrafted features with deep features to classify each nodule as either benign or malignant, achieving an AUC of 0.96. Alakwaa et al. [12] combined the LUNA16 [13] dataset with a subset of the National Lung Screening Trial (NLST) [14], then used a pre-trained U-Net to segment potential nodules from a CT scan automatically. The segmented nodules were passed to a 3D CNN to detect early-stage lung cancer, achieving an AUC of 0.83 in a randomly-split test cohort from the abovementioned data. Ardila et al. [15] developed an end-to-end set of 3D CNN modules to compute the overall risk of lung malignancy based on autodetection of nodules, using the full-size publicly available NLST dataset. In a retrospective reader study, their model outperformed six experienced radiologists with absolute reductions of 11% and 5% in false positives and false negatives, respectively.

The need for transparency, interpretability and explainability in such computer-aided diagnostic recommendations will grow to become increasingly prominent in the immediate future. A crucial piece of law, the General Data Protection Regulation (GDPR), governs the rights of European Union (EU) citizens as human data subjects and addresses processing by automated means for decision-making anywhere in the world if it concerns an EU individual. Specifically, the GDPR enshrines the right of an individual to receive "meaningful information about the logic involved" in an automated decision concerning them, and on that basis to either legally challenge the decision, or exercise conscientious objection to the use of an automated means for deriving the decision [16].

While definition of "meaningful" is open for debate, it is clearly helpful to be able to point at specific regions of interest (ROIs) that were strongly triggering for the diagnostic recommendation, along with related features of lung cancer and non-lung cancer cases. In this way, a human radiologist can review the information in depth, and either confirm or over-rule the recommendations of an automated system. Irrespective of a right to an explanation, a computerized diagnostic support system with high transparency and high interpretability would be immensely valuable in clinical practice.

For automated diagnosis of lung cancer, a deep learning-based system can be applied in two levels: at nodule level, to identify potential malignant nodule(s) for further biopsy and performing diagnosis at patient level. Generally speaking, nodule classification methods need a label for each nodule to be able to train a model [9][11]. However, labelling each nodule is more time-consuming and expensive than having a label for each patient, which is usually already available in hospital records. In this study we focus on deep learning methods for lung cancer diagnosis that can make use of the existing data to develop a lung cancer CAD system that classifies patients based on multiple suspected nodules in the entire CT series without the need to assign a label to each nodule (i.e. each instance), and at the same time provide high visibility of the triggering features of its recommendation. Multiple instance learning (MIL) with attention mechanism [17][19] fits this need well. In MIL, the nodules are grouped into 'bags of instances' (assuming multiple nodules in one CT examination of the chest area). The task is hence to determine the diagnosis for the subject as a whole. Only the subject-level diagnoses (i.e. the bag labels) are needed, but not individual labels of every nodule found in the subject [20]. This approach is thus more amenable to real-world data mining in lung cancer, since the subject level diagnosis is much more widely available than annotations on each nodule.

Research on MIL problems has progressed along instance-level versus embedding-level solutions [21], with the latter seeming to perform better at subject-level classification [22]. Widely-used embedding approaches include MI-SVM [23], mi-Graph [24], miVLAD [25] and MI-Net [25], but the shortcoming of these is the lack of transparency of triggering instance(s). An attention-based deep MIL [21][33] has been recently introduced, that allows a deep learning model to estimate the contribution of each instance to the predicted subject label, using the well-established attention mechanism [26].

The objective of this work was to develop a lung cancer classification model at the subject (patient) level from multiple examined nodules, without the need to have specific expert findings reported at the level of each individual nodule. An MIL method with an additional deep attention mechanism was used to help draw an expert clinician's eye towards the individual nodules that were strongly triggering for the model's diagnostic recommendation. We propose that this will be important by way of offering better interpretability and the possibility of human expert verification of the internal logic of the algorithm. A selection of commonly-used hand-crafted radiomics features was used as a source of image features, and we also compared a number of alternative MIL methods. We have re-used an existing open access data collection for training and cross-validation.

This article is organized as follows: the methods and classifier experiments are described in Section 2. Our results are given in Section 3. The significance of our findings and limitations of our current approach will be discussed in Section 4. An overall summary and conclusion are presented in Section 5. Finally, source code will open access for public at https://gitlab.com/UM-CDS/combine-mil-and-radiomics-for-lung-screening) and additional details of the system architecture will be given in Supplementary Materials.

**2. Methods**

*2.1 Dataset*

The primary data source of data is an open access collection from the Lung Image Database Consortium (LIDC-IDRI) [31], accessed at The Cancer Imaging Archive (TCIA) during May 2020 [32] under a Creative Commons Attribution Non-Commercial 3.0 Unported (CC BY-NC) license. The details of subjects in LIDC-IDRI have been provided elsewhere [31], but briefly: the collection comprises 1018 clinical chest CT examinations from seven disjoint institutions. Radiologists working independently entered 7371 annotations, of which there were 2669 consensus nodules. We excluded subjects with unreported or unknown diagnosis, and excluded nodules below 3mm in diameter according to current diagnosis protocols [34][35]. This resulted in 110 unique subjects with a total of 310 nodules eligible for consideration. Binary masks for the nodules were provided in the data collection as an XML file. Numbers of subjects and nodules excluded, along with reason, are provided in Figure 1 below. From the summary of diagnostic findings in Table 1, we note that the majority of subjects and lung nodules in the dataset are positive for lung cancer; 75% and 77%, respectively. Index of available patients for experiments in LIDC-IDRI can be found in Supplementary Table 1.

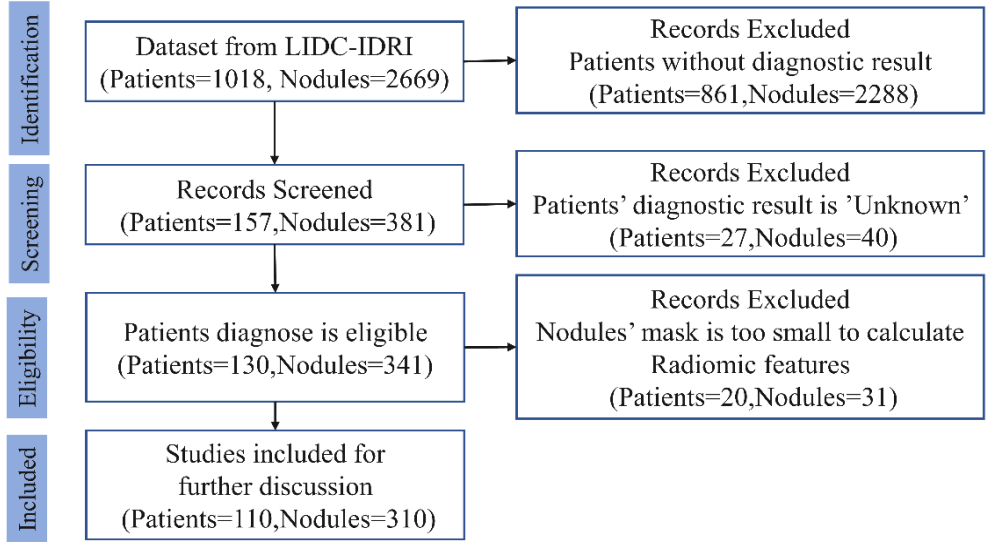

Figure 1. Sample selection flowchart describing the number of subjects and the number of nodules selected for this analysis.

Table 1 summarizes the radiological findings available in the selected subset with definitive subject-level diagnosis and nodule-level classification.

Table 1. Number of patients and nodules according to ground truth diagnosis in the dataset

|  | **Lung cancer** | **Not lung cancer** | Total |
|---|---|---|---|
| Numbers of (% of total) patients | 82 (75%) | 28 (25%) | 110 |
| Numbers of (% of total) nodules | 239 (77%) | 71 (23%) | 310 |

*2.2 Image acquisition settings*

The LIDC-IDRI contains a heterogeneous set of CT of subjects from different institutions. We used axial CT images with dimension of 512x512 pixels. Radiation exposure of selected samples ranged from 3 milliampere-seconds (mAs) to 534 mAs (median: 147.5 mAs), and reconstructed slice thicknesses ranging from 0.6 mm to 5.0 mm (median: 2.0 mm).

*2.3 Feature extraction*

Radiomics features were extracted using an open-source Python library pyRadiomics (v2.2.0) [36]. Images were resampled to 2 mm isotropic voxels prior to feature extraction. A total of 103 features were extracted. These consisted of 13 morphology (shape) features, 17 intensity-histogram (first-order) features and 73 textural (Haralick) features. Binary masks for the GTV were generated from the XML file in the LIDC-IDRI collection, using an open-access library *pylidc* [37]. DICOM CT images were converted to 3D images by using SimpleITK (v1.2.4) [38] for pyRadiomics feature extraction. The mathematical definition of each feature has been given in the online documentation. Our pyRadiomics extraction settings (from the params.yaml file) have been included in Supplementary Table 2. All features included in this analysis have been listed in Supplementary Table 3.

*2.3 Classifier*

We used an attention-based MIL for the lung cancer classifier component. This consists of two parts that can be trained end-to-end. First, the transformation network was implemented as three fully connected neural network layers with a dropout rate of 0.5. Additional details about this network are in Supplementary Table 4. To fix the dimension of the input layer of neurons, the 103 features per nodule were duplicated within the same subject until it was the same as the maximum number of nodules per subject, which we found to be 12 in this case. More specifically, each nodule in the same bag should be duplicated with the same probability. For example, if there are 5 nodules in a bag, 3 random nodules need to be duplicates once (i.e., appear twice in total) and 2 random nodules need to be duplicated twice (appear 3 times in total) in the final fixed feature bags. Therefore, the dimension of the input layer should be 103 and one bag consists of 12 vectors (103 x 12). Feature duplication was performed before model training and was also used in model testing.

Second, the attention-based pooling layer implemented the attention mechanism popularized by long short-term memory networks (LSTMs) [39]. The attention mechanism is an important strategy that fits encoder input sequences into a fixed-length internal representation. The architecture of the classifier is illustrated schematically in Figure 2.

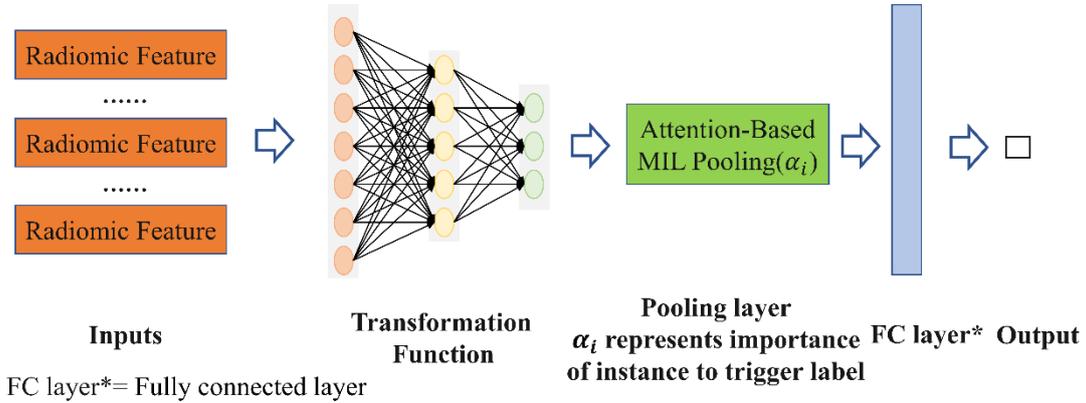

Figure 2. Architecture of the Attention-based Deep MIL. Extracted radiomics features are used as the input to the transformation network, which is then pooled with attention. A fully-connected final layer combines the attention-based pooling to give the output probability.

*2.4 Addressing class imbalance*

Imbalance in the outcome frequency (i.e. lung cancer versus not lung cancer) has been known to affect the classifier, biasing this towards the dominant class. Several methods are available to address class imbalance [41] in general, and we applied a novel sampling method to address class imbalance specifically for MIL. It is assumed that all nodules in non-cancer subjects are, by clinical definition, non-cancerous nodules. Synthetic non-cancer patients were thus generated by randomly sampling a finite number of instances out of all the nodules in an aggregated pool of actual non-cancer subjects. On the other hand, synthetic cancer patients could be generated by adding a random number of negative instances sampled from the instances pool (from both negative and positive bags) to the original positive bags. However, we did not simulate cancer patients in our experiments, because positive bags were majority in our dataset. This was only done for the training set; no class imbalance correction was applied in the testing set.

*2.5 Model development and validation*

All work was executed on a Core i7 8565U CPU with 8GB of RAM. The optimizer for network training was stochastic gradient descent (SGD) [42], with batch size 1 and the learning rate fixed at 0.0001. The neural network was trained for 500 epochs (taking 3-4 minutes) per experiment.

We performed experiments for the attention-based MIL in comparison with other MIL approaches - MI-SVM, mi-graph, miVLAD, MI-Net and a naïve MIL algorithm that performs a simple aggregation of the predictions by replacing the attention-based MIL pooling with average MIL pooling [20]. The optimizer, batch size, learning rate and training epochs were set same as attention-based MIL in MI-Net. The setting of hyperparameters in other methods were followed as mentioned in original literatures [23][24][25]. Same training and testing data were used in every running for all methods.

Model training was performed on all the available subjects, taking their respective diagnosis as the "bag label" and the nodules as the instances. We ran 20 repetitions of end-to-end training runs on the hand-crafted features with 5-fold cross-validation in each run and there is no oversampling in testing dataset. For each repetition of 5-fold cross-validation, samples were randomly sorted first and then split into 5 folds, so that each sample was used once for testing and 4 times for training in each repetition. We adjusted for the lower number of non-cancer diagnoses by generating synthetic non-cancer patients as described above (section 2.4). Specifically, we synthesized 60 additional non-cancer subjects from the initial training dataset and added these to the actual 88 training subjects, resulting in a training set containing 148 subjects in total. No synthetic re-sampling was used for positive lung cancer subjects. We further conducted an additional sensitivity analysis to assess how oversampling to overcome class imbalance might have affected the model's performance by using only the original data of 110 subjects.

The discriminative performance was assessed using the mean and standard error of the mean (SEM) of recall, accuracy, positive predictive value (PPV), negative predictive value (NPV), respectively. For dichotomization of outcome, we used a probability threshold of 0.5 to separate lung cancer from non-lung cancer. The area under the receiver operating characteristic curve (AUC) was computed for each model, the definition of AUC can be found in [43]. Let TP, TN, FP, and FN denote true positive, true negative, false positive, and false negative, respectively, then we define recall, accuracy, PPV and NPV as:

$$recall = \frac{TP}{TP+TN} \qquad (1)$$

$$accuracy = \frac{TP+TN}{TP+TN+FR+FN} \qquad (2)$$

$$PPV = \frac{TP}{TP+FP} \qquad (3)$$

$$NPV = \frac{TN}{TN+FN} \qquad (4)$$

All statistical analysis was done in Python (version 3.6.1).

## 3. Results

Figure 3 shows the violin plots comparing the results of attention-based MIL with (3a) and without (3b) synthetic minority oversampling. The estimated mean (with SEM in the parentheses) for recall, accuracy, PPV, NPV and the AUC for the model including the class imbalance correction were: 0.870 (SEM 0.061), 0.807 (SEM 0.069), 0.928 (SEM 0.078), 0.591 (SEM 0.155) and 0.842 (SEM 0.071) respectively. Without the class imbalance correction, these values were: 0.889 (SEM 0.061), 0.768 (SEM 0.059), 0.842 (SEM 0.071), 0.483 (SEM 0.209) and 0.696 (SEM 0.108) respectively. The main effect of the minority oversampling was to improve accuracy, PPV, NPV and AUC. A representative (from a selected repetition) set of AUC curves for the different MIL methods with the same training and testing data can be found in Figure 4.

Table 2 summarizes the results of comparing different MIL approaches. Attention based MIL without oversampling achieved the best recall, MI-Net achieved the best PPV and attention based MIL achieved the best accuracy, PPV and AUC. Attention based MIL was better than other methods in PPV and AUC significantly (Wilcoxon test, $p <0.01$), however, attention based MIL was worse than best result in recall and NPV (Wilcoxon test, $p = 0.02$ and $p <0.01$ respectively). Moreover, attention based MIL with oversampling is better than attention-based MIL without oversampling in all metrics significantly except recall (Wilcoxon test, $p <0.01$ for accuracy, PPV, NPV and AUC, $p =0.02$ for recall).

The absence of AUCs for mi-graph and miVLAD is due to our reusing of the source code by the LAMDA lab, Nanjing University [44]. Their source code for mi-graph and miVLAD outputs only the classification label (not the probability) and therefore, the AUCs cannot be calculated.

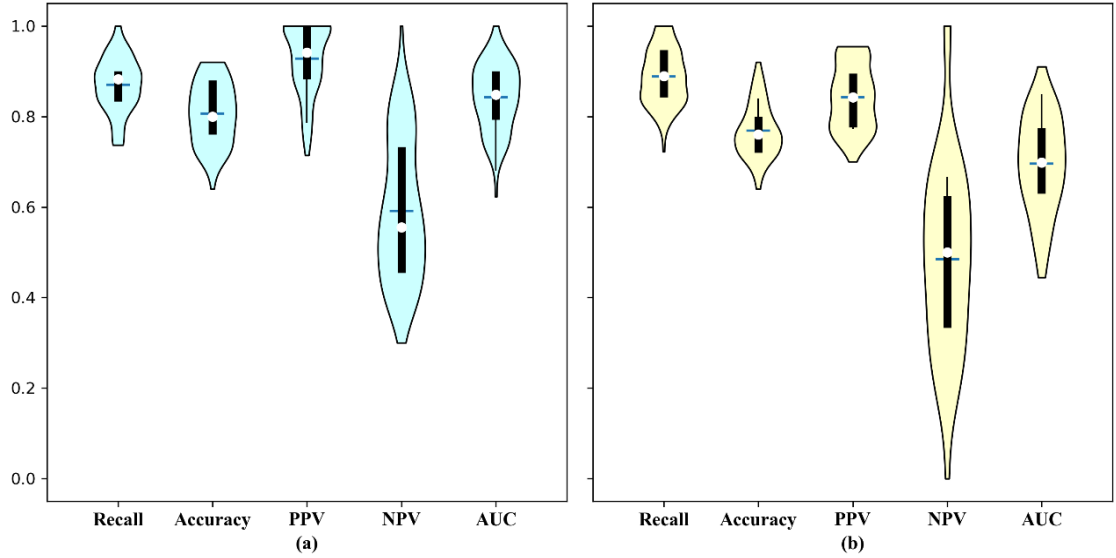

Figure 3. Violin plot of the experimental results (a) with oversampling and (b) without oversampling.

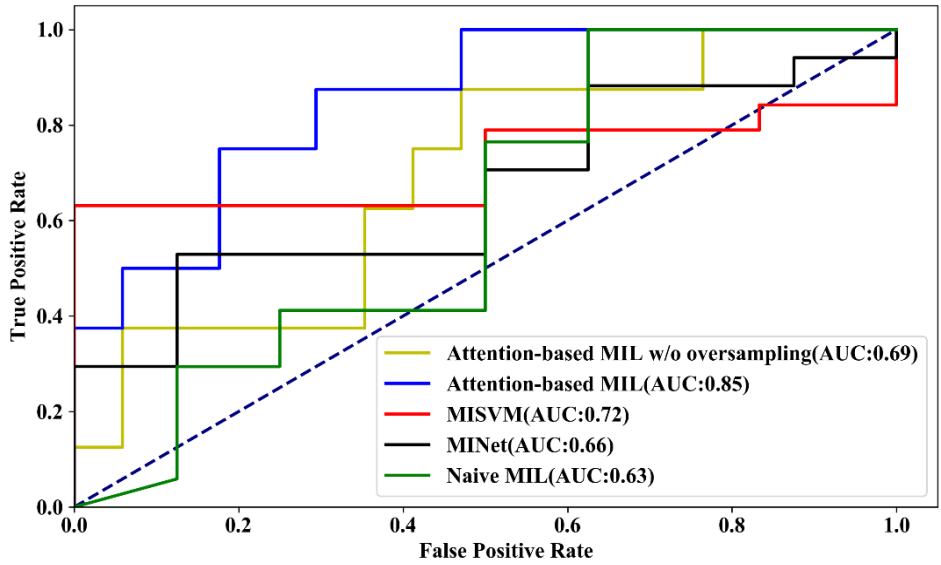

Figure 4. An example of AUC curves for different methods with same training and testing data. An AUC curves for Attention-based MIL, Attention-based MIL w/o oversampling, MI-SVM, MI-Net and Naïve MIL.

Table 2. Results of the Attention based Deep MIL approach with class imbalance correction, compared to other MIL methods (Attention-based MIL w/o oversampling, MI-SVM, mi-graph, miVLAD and MI-Net)

| Methods | Attention-based MIL | Attention-based MIL w/o oversampling | MI-SVM | mi-graph | miVLAD | MI-Net | Traditional MIL |
|---|---|---|---|---|---|---|---|
| Recall | $0.870 \pm 0.061$ | $\mathbf{0.889 \pm 0.061}$ | $0.756 \pm 0.084$ | $0.777 \pm 0.048$ | $0.871 \pm 0.087$ | $0.835 \pm 0.109$ | $0.850 \pm 0.099$ |
| Accuracy | $\mathbf{0.807 \pm 0.069}$ | $0.768 \pm 0.059$ | $0.703 \pm 0.080$ | $0.749 \pm 0.055$ | $0.782 \pm 0.063$ | $0.727 \pm 0.050$ | $0.748 \pm 0.065$ |
| PPV | $\mathbf{0.928 \pm 0.078}$ | $0.842 \pm 0.071$ | $0.560 \pm 0.199$ | $0.772 \pm 0.042$ | $0.835 \pm 0.059$ | $0.522 \pm 0.265$ | $0.835 \pm 0.070$ |
| NPV | $0.591 \pm 0.155$ | $0.483 \pm 0.209$ | $0.810 \pm 0.080$ | $0.713 \pm 0.229$ | $0.675 \pm 0.160$ | $\mathbf{0.838 \pm 0.069}$ | $0.478 \pm 0.233$ |
| AUC | $\mathbf{0.842 \pm 0.071}$ | $0.696 \pm 0.108$ | $0.625 \pm 0.099$ | -- | -- | $0.662 \pm 0.093$ | $0.681 \pm 0.080$ |

In order to determine the level of oversampling, we ran sensitivity analyses. We gradually increased the number of included simulated non-cancer subjects from 0 to 100 on steps of 20. We ran 20-repeat 5-fold cross-validation for each experiment. The results of sensitivity analysis are shown in Figure 5.

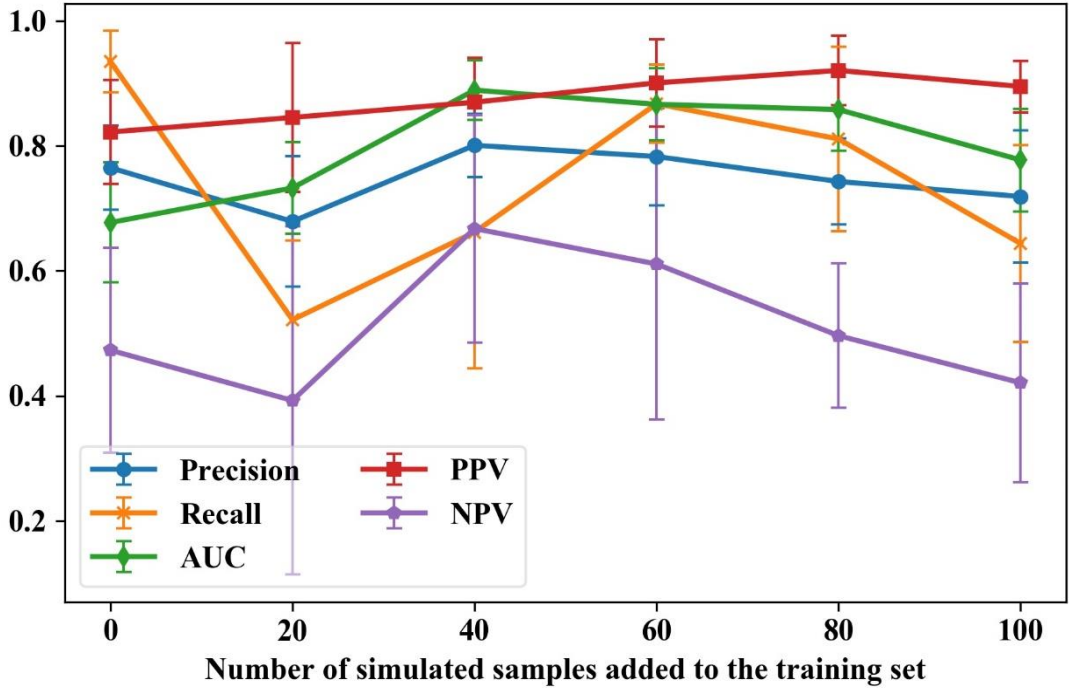

Figure 5. Results of sensitivity analysis for different levels of oversampling.

As shown in Figure 5, including 60 simulation samples results in good performance for all metrics (especially for Recall) with less computation compared with other settings with similar performance.

Given how important batch size is for convolutional neural networks training [45], we ran a sensitivity analysis on this parameter. We ran 20-repeat 5-fold cross-validation analyses with different batch sizes (1, 2, 3 and 4) for each experiment. The loss curves for model training with different batch sizes is shown in Figure 6 (a) and the performance of models trained with different batch sizes is shown in Figure 6(b).

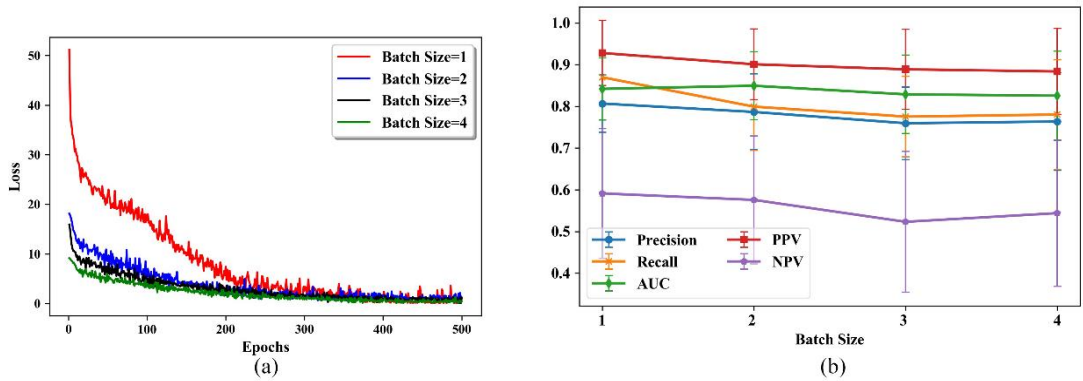

Figure 6. Results of sensitivity analysis for different batch sizes. (a) Loss curves for model training with different batch size; (b) performance of models trained with different batch sizes.

As shown in Figure 6, the model trained with a batch size of 1 achieved the best performance according to all metrics except AUC (0.842 for batch size 1 vs 0.849 for batch size 2) and the loss of all models converged at the end of the 500 epochs. Therefore, we set the batch size to 1 in this study.

Besides model performance, one of the most appealing aspects that we selected the attention-based

MIL method for, was to indicate the instances that might have been strongly influential on the classification. In this case, it would be the relative importance of each nodule when predicting the subject label as either lung cancer or not lung cancer. A couple of lung cancer examples are shown in Figure 7 for two subjects in the dataset, LIDC-IDRI-1004 and LIDC-IDRI-1011.

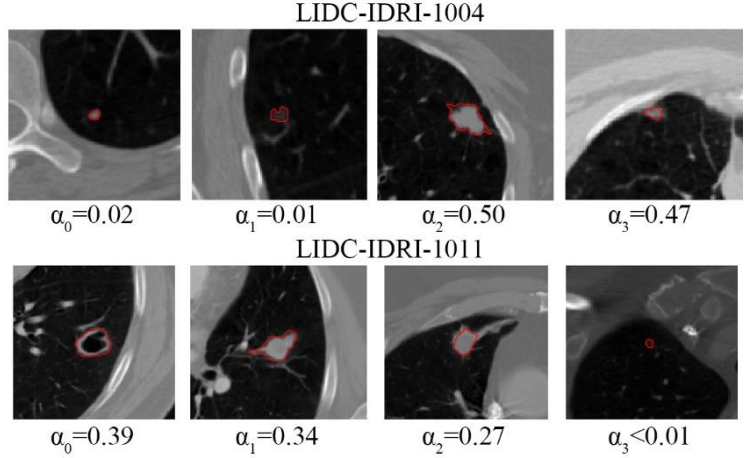

Figure 7. An example of attention weights for two positive lung cancer subjects (LIDC-IDRI-1004 and 1011).

Alpha in Figure 7 mean the strength of the attention, value of alpha only meaningful in the same patient and it is meaningless by comparing alphas across patients. The order of nodules was arranged in random way within same patient.

The evaluation of the attention mechanism was performed by one of co-authors -- a radiologist with 3-year experience, who examined some sample patients' weights and agreed with the weighting. In these examples, it is clearly discernable from the weights ($\alpha_2$ and $\alpha_3$ larger than $\alpha_0$ and $\alpha_1$) that the two rightmost nodules pictured for subject LIDC-IDRI-1004 are much more strongly influential in the diagnostic evaluation compared to the two leftmost nodules. Similarly, for subject LIDC-IDRI-1011, three of the nodules are influential on the subject classification, but the nodule pictured rightmost is not influential at all (alpha < 0.01).

## 4. Discussion

Our objective was to propose a lung cancer classification at the subject level from multiple examined nodules, with an attention mechanism for improving the interpretability. The results show that our proposed classification achieves good performance compared to other MIL methods and that the unique characteristic of the deep attention-based MIL, namely attention weights, potentially makes our method more interpretable for clinicians.

To see the effect of minority oversampling to overcome class imbalance, we tested the model with and without the oversampling. The results show that the oversampling improved the model's performance significantly in accuracy, PPV, NPV and AUC by comparing Attention-based MIL without oversampling. However, there are seem some decrease in recall.

We observed from Figure 3 that minority oversampling has a major effect on the AUC. In fact, the AUC sinks below 0.5 in some experiments without oversampling. This can be explained by the fact that the AUC is more sensitive to the classification performance of the model with the minority class than either accuracy or recall [40].

We proposed a new synthetic subject generation method that can be used to overcome class imbalance by oversampling the minority class. We did this by sampling from an aggregated pool of nodules from patients with the ground truth of "not lung cancer". To the best of our knowledge, such methods have not yet been proposed for MIL. This oversampling technique resulted in significant improvements on accuracy, PPV, NPV and AUC. We believe this strategy, which is based on the characteristics of MIL, can be used when training any MIL model from a class imbalanced dataset.

The results show that our method could potentially be applied to automated lung cancer diagnosis, subject to further validation and studies in large datasets. However, we acknowledge there are some limitations and weaknesses in the assumptions we had to make. First, due to the need of a mask that delineates the nodules to calculate radiomic features, our method would have to be dependent on lung nodule detection and segmentation methods such as the ones proposed by Huang et al. [46] and Anirudh et al. [47]. This dependence on pre-existing or human expert segmentation is not new, and is problem that still affects many aspects of medical image analysis and supervised machine learning. Related to this is the potential for inter-observer disagreement about the external outline of the nodule. This problem is well known and documented for large and locally-advanced lung tumours, but with the small nodule volumes involved in this study, we have assumed that the inter-observer problem does not strongly affect the extracted features. A further question we cannot address in this study is the problem of undetected nodules and very small nodules (diameter smaller than 3mm) that were omitted, moreover, images were resampled to 2 mm isotropic voxels prior to feature extraction, which is possibly also a reason why very small nodules are not appropriate. This work has assumed no false positives and no false negatives, so we cannot elucidate what happens with imperfect nodule detection.

The performance of our model appears sensitive to sampling effects, in other words, the performance of model fluctuates across repeated experiments, as shown in Figure 3. This is likely a direct consequence of the relatively small sample size of the dataset. Expanding the sample size by including small nodules is not immediately helpful because they do not add that many subjects and nodules to the sample, whereas hand-crafted features would not be stable when taken from very small volumes. The major root of the problem appears to be the lack of ground truth and annotated images. Related to this fact is that we currently did not find a suitable dataset for external, independent validation. Therefore, our results should be interpreted as preliminary indication of feasibility, and larger datasets need to be used to demonstrate wider generalizability of this work.

Due to the high fitting ability of neural networks and large epochs during training, the model returns 1 or 0 almost all of the time, which means the overall model calibration was generally poor [48][49]. Model calibration plot is shown in Supplementary Figure 1, and it appears that all MIL methods have poor calibration except MI-SVM.

In addition, we have not explored feature dimensionality reduction and applied feature redundancy analysis. This is in part due in part to the transformation network that does not require explicit feature selection steps prior to MIL pooling. The repeatability and reproducibility of handcrafted features are subjects of numerous investigations in radiomics and appears to be highly modality specific. This work has not explored the stability of low-dose CT-derived image features, which tend to have quite a lot of noise present [50]. This could affect the performance of our model in an external validation, and image harmonization or denoising strategies may be needed in future to support general extensibility.

Moreover, we were not able to test the performance of the models in an external dataset, which would have provided more reliable estimates of the models' potential performance in a different setting . On the other hand, the dataset used in this study (LIDC-IDRI), was collected over 10 years ago. With new emerging CT technologies and reconstruction methods, it is possible that different conclusions would be reached if the proposed method is applied to newer images currently being used in clinical practice. Further research on this aspect is required.

Finally, our oversampling strategy is sensitive to the quality of data's label at patient level. More specifically, if labels are incorrect (e.g. if one or two of the nodules has been misclassified by error and the subject is hence a false negative), the noise will be amplified due to oversampling.

For future work, an automated nodule detection and segmentation algorithm could be attached to this attention-based MIL classifier to fully complete the lung cancer diagnosis workflow. Secondly, methods for improving radiomic features' reliability in low dose CT may be necessary for improving model's performance in unseen data. Thirdly, large scale and comprehensive evaluation of the attention mechanism is needed in the future to assess its reliability and reproducibility. Fourthly, a comparison between the proposed method and a traditional deep learning-based image classification algorithm would be of special interest. Finally, the proposed model needs to be externally validated to assess whether the model suffers from overfitting to the training data or whether it is widely generalizable to CT images from different scanners.

## 5. Conclusion

We treated computer-aided diagnosis of lung cancer as a multiple instance learning (MIL) problem, such that the classification as lung cancer or not is made at the subject level (i.e. the patient) without relying on classifications at the level of individual nodules (i.e. each of suspicious lung nodules). The addition of the attention mechanism was used to draw the clinician's eye towards features that were important for triggering the recommended diagnosis, with the aim of supporting interpretability and, importantly, verification by human experts of the algorithm's internal logic. We used radiomics as a source of interpretable image-derived features, and deep attention-based MIL was found to be a superior classifier compared to other MIL options with regard to accuracy, NPV and AUC. A novel approach for minority oversampling, adapted for MIL problems, has been used to address the outcome class imbalance in the LIDC-IDRI dataset. We showed how an attention mechanism could be used as an indication of the importance of each nodule for triggering the diagnostic recommendation. Cross-validation was used to check for model performance, but more data is required to provide a robust test of wider generalizability.

## 6. Acknowledgements

The authors would like to thank the authors of the open-access source code for mi-graph and miVLAD from LAMDA lab（http://210.28.132.67/Data.ashx). JC is supported by a China Scholarship Council scholarship (201906540036). HZ is supported by a China Scholarship Council scholar (201909370087). The co-authors acknowledge funding support from the following: STRaTegy (STW 14930), BIONIC (NWO 629.002.205), TRAIN (NWO 629.002.212), CARRIER (NWO 628.011.212) and The Hanarth Foundation for LW.

## 7. Reference

[1] Cancer. (2018) [WWW document]. URL https://www.who.int/news-room/fact-sheets/detail/cancer. [access on 20 June 2020]

[2] Siegel, R. L., Miller, K. D., & Jemal, A. (2020). Cancer statistics, 2020. *CA: a cancer journal*

*for clinicians*, *70*(1), 7–30. https://doi.org/10.3322/caac.21590

[3] National Lung Screening Trial Research Team. (2011). Reduced lung-cancer mortality with low-dose computed tomographic screening. *New England Journal of Medicine*, *365*(5), 395-409.

[4] Kanazawa, K., Kawata, Y., Niki, N., Satoh, H., Ohmatsu, H., Kakinuma, R., ... & Eguchi, K. (1998). Computer-aided diagnosis for pulmonary nodules based on helical CT images. *Computerized medical imaging and graphics*, *22*(2), 157-167.

[5] Naqi, S. M., Sharif, M., & Jaffar, A. (2020). Lung nodule detection and classification based on geometric fit in parametric form and deep learning. *Neural Computing and Applications*, *32*(9), 4629-4647.

[6] Parveen, S. S., & Kavitha, C. (2012). A review on computer aided detection and diagnosis of lung cancer nodules. *International Journal of Computers & Technology*, *3*(3a), 393-400.

[7] Yang, Y., Feng, X., Chi, W., Li, Z., Duan, W., Liu, H., ... & Liu, B. (2018). Deep learning aided decision support for pulmonary nodules diagnosing: a review. *Journal of thoracic disease*, *10*(Suppl 7), S867.

[8] Krizhevsky, A., Sutskever, I., & Hinton, G. E. (2012). Imagenet classification with deep convolutional neural networks. In *Advances in neural information processing systems* (pp. 1097-1105).

[9] Shen, W., Zhou, M., Yang, F., Yang, C., & Tian, J. (2015, June). Multi-scale convolutional neural networks for lung nodule classification. In *International Conference on Information Processing in Medical Imaging* (pp. 588-599). Springer, Cham.

[10] Zhou, B., Khosla, A., Lapedriza, A., Oliva, A., & Torralba, A. (2016). Learning deep features for discriminative localization. In *Proceedings of the IEEE conference on computer vision and pattern recognition* (pp. 2921-2929).

[11] Xie, Y., Zhang, J., Xia, Y., Fulham, M., & Zhang, Y. (2018). Fusing texture, shape and deep model-learned information at decision level for automated classification of lung nodules on chest CT. *Information Fusion*, 42, 102-110.

[12] Alakwaa, W., Nassef, M., & Badr, A. (2017). Lung cancer detection and classification with 3D convolutional neural network (3D-CNN). *Lung Cancer*, 8(8), 409.

[13] Setio, A. A. A., Traverso, A., De Bel, T., Berens, M. S., van den Bogaard, C., Cerello, P., ... & van der Gugten, R. (2017). Validation, comparison, and combination of algorithms for automatic detection of pulmonary nodules in computed tomography images: the LUNA16 challenge. *Medical image analysis*, *42*, 1-13.

[14] Black, W. C., Gareen, I. F., Soneji, S. S., Sicks, J. D., Keeler, E. B., Aberle, D. R., ... & Gatsonis, C. (2014). Cost-effectiveness of CT screening in the National Lung Screening Trial. *The New England journal of medicine*, *371*(19), 1793-1802.

[15] Ardila, D., Kiraly, A. P., Bharadwaj, S., Choi, B., Reicher, J. J., Peng, L., ... & Naidich, D. P. (2019). End-to-end lung cancer screening with three-dimensional deep learning on low-dose chest computed tomography. *Nature medicine*, *25*(6), 954-961.

[16] Selbst, A., & Powles, J. (2018, January). "Meaningful Information" and the Right to

Explanation. In *Conference on Fairness, Accountability and Transparency* (pp. 48-48). PMLR.

[17] Data protection in the EU. (2020) [WWW document]. URL https://ec.europa.eu/info/law/law-topic/data-protection/data-protection-eu_en. [access on 5 July 2020]

[18] Dietterich, T. G., Lathrop, R. H., & Lozano-Pérez, T. (1997). Solving the multiple instance problem with axis-parallel rectangles. *Artificial intelligence*, *89*(1-2), 31-71.

[19] Maron, O., & Lozano-Pérez, T. (1998). A framework for multiple-instance learning. In *Advances in neural information processing systems* (pp. 570-576).

[20] Bhattacharjee, K., Pant, M., Zhang, Y. D., & Satapathy, S. C. (2020). Multiple Instance Learning with Genetic Pooling for medical data analysis. *Pattern Recognition Letters*, *133*, 247-255.

[21] Ilse, M., Tomczak, J. M., & Welling, M. (2018, January). Attention-based deep multiple instance learning. In *35th International Conference on Machine Learning, ICML 2018* (pp. 3376-3391). International Machine Learning Society (IMLS).

[22] Wang, X., Yan, Y., Tang, P., Bai, X., & Liu, W. (2018). Revisiting multiple instance neural networks. *Pattern Recognition*, *74*, 15-24.

[23] Andrews, S., Tsochantaridis, I., & Hofmann, T. (2003). Support vector machines for multiple-instance learning. In *Advances in neural information processing systems* (pp. 577-584).

[24] Zhou, Z. H., Sun, Y. Y., & Li, Y. F. (2009, June). Multi-instance learning by treating instances as non-iid samples. In *Proceedings of the 26th annual international conference on machine learning* (pp. 1249-1256).

[25] Wei, X. S., Wu, J., & Zhou, Z. H. (2016). Scalable algorithms for multi-instance learning. *IEEE transactions on neural networks and learning systems*, *28*(4), 975-987.

[26] Sutskever, I., Vinyals, O., & Le, Q. V. (2014). Sequence to sequence learning with neural networks. In *Advances in neural information processing systems* (pp. 3104-3112).

[27] Afshar, P., Mohammadi, A., Plataniotis, K. N., Oikonomou, A., & Benali, H. (2019). From handcrafted to deep-learning-based cancer radiomics: challenges and opportunities. IEEE Signal Processing Magazine, 36(4), 132-160.

[28] Aerts, H. J., Velazquez, E. R., Leijenaar, R. T., Parmar, C., Grossmann, P., Carvalho, S., ... & Hoebers, F. (2014). Decoding tumour phenotype by noninvasive imaging using a quantitative radiomics approach. *Nature communications*, *5*(1), 1-9.

[29] Lambin, P., Rios-Velazquez, E., Leijenaar, R., Carvalho, S., Van Stiphout, R. G., Granton, P., ... & Aerts, H. J. (2012). Radiomics: extracting more information from medical images using advanced feature analysis. *European journal of cancer*, *48*(4), 441-446.

[30] Susto, G. A., Schirru, A., Pampuri, S., McLoone, S., & Beghi, A. (2014). Machine learning for predictive maintenance: A multiple classifier approach. *IEEE Transactions on Industrial Informatics*, *11*(3), 812-820.

[31] Armato III, S. G., McLennan, G., Bidaut, L., McNitt-Gray, M. F., Meyer, C. R., Reeves, A. P., ... & Kazerooni, E. A. (2011). The lung image database consortium (LIDC) and image database resource initiative (IDRI): a completed reference database of lung nodules on CT scans. *Medical physics*, *38*(2), 915-931.

[32] Clark, K., Vendt, B., Smith, K., Freymann, J., Kirby, J., Koppel, P., ... & Tarbox, L. (2013). The

Cancer Imaging Archive (TCIA): maintaining and operating a public information repository. *Journal of digital imaging*, *26*(6), 1045-1057.

[33] Han, Z., Wei, B., Hong, Y., Li, T., Cong, J., Zhu, X., ... & Zhang, W. (2020). Accurate Screening of COVID-19 using Attention Based Deep 3D Multiple Instance Learning. *IEEE Transactions on Medical Imaging*.

[34] Han, F., Wang, H., Zhang, G., Han, H., Song, B., Li, L., ... & Liang, Z. (2015). Texture feature analysis for computer-aided diagnosis on pulmonary nodules. *Journal of digital imaging*, *28*(1), 99-115.

[35] Dhara, A. K., Mukhopadhyay, S., Dutta, A., Garg, M., & Khandelwal, N. (2016). A combination of shape and texture features for classification of pulmonary nodules in lung CT images. *Journal of digital imaging*, *29*(4), 466-475.

[36] Van Griethuysen, J. J., Fedorov, A., Parmar, C., Hosny, A., Aucoin, N., Narayan, V., ... & Aerts, H. J. (2017). Computational radiomics system to decode the radiographic phenotype. *Cancer research*, *77*(21), e104-e107.

[37] Hancock, M. C., & Magnan, J. F. (2016). Lung nodule malignancy classification using only radiologist-quantified image features as inputs to statistical learning algorithms: probing the Lung Image Database Consortium dataset with two statistical learning methods. *Journal of Medical Imaging*, *3*(4), 044504.

[38] Yaniv, Z., Lowekamp, B. C., Johnson, H. J., & Beare, R. (2018). SimpleITK image-analysis notebooks: a collaborative environment for education and reproducible research. *Journal of digital imaging*, *31*(3), 290-303.

[39] Hochreiter, S., & Schmidhuber, J. (1997). Long short-term memory. *Neural computation*, *9*(8), 1735-1780.

[40] He, H., & Garcia, E. A. (2009). Learning from imbalanced data. *IEEE Transactions on knowledge and data engineering*, *21*(9), 1263-1284.

[41] Haixiang, G., Yijing, L., Shang, J., Mingyun, G., Yuanyue, H., & Bing, G. (2017). Learning from class-imbalanced data: Review of methods and applications. *Expert Systems with Applications*, *73*, 220-239.

[42] Bottou, L. (2012). Stochastic gradient descent tricks. In *Neural networks: Tricks of the trade* (pp. 421-436). Springer, Berlin, Heidelberg.

[43] Huang, Jin, and Charles X. Ling. "Using AUC and accuracy in evaluating learning algorithms." *IEEE Transactions on knowledge and Data Engineering* 17.3 (2005): 299-310.

[44] Learning and Mining from DatA. (2021) [WWW document]. URL http://210.28.132.67/Data.ashx. [accesses on 4 Jun 2021]

[45] Yong, H., Huang, J., Meng, D., Hua, X., & Zhang, L. (2020, August). Momentum batch normalization for deep learning with small batch size. In *European Conference on Computer Vision* (pp. 224-240). Springer, Cham.

[46] Huang, X., Shan, J., & Vaidya, V. (2017, April). Lung nodule detection in CT using 3D convolutional neural networks. In *2017 IEEE 14th International Symposium on Biomedical Imaging (ISBI 2017)* (pp. 379-383). IEEE.

[47] Anirudh, R., Thiagarajan, J. J., Bremer, T., & Kim, H. (2016, March). Lung nodule detection using 3D convolutional neural networks trained on weakly labeled data. In *Medical Imaging 2016: Computer-Aided Diagnosis* (Vol. 9785, p. 978532). International Society for Optics and Photonics.

[48] Steyerberg, E. W., Vickers, A. J., Cook, N. R., Gerds, T., Gonen, M., Obuchowski, N., ... & Kattan, M. W. (2010). Assessing the performance of prediction models: a framework for some traditional and novel measures. *Epidemiology (Cambridge, Mass.)*, *21*(1), 128.

[49] Steyerberg, E. W., & Vergouwe, Y. (2014). Towards better clinical prediction models: seven steps for development and an ABCD for validation. *European heart journal*, *35*(29), 1925-1931.

[50] Bagher-Ebadian, H., Siddiqui, F., Liu, C., Movsas, B., & Chetty, I. J. (2017). On the impact of smoothing and noise on robustness of CT and CBCT radiomics features for patients with head and neck cancers. *Medical physics*, *44*(5), 1755-1770.

[51] Moons, K. G., Altman, D. G., Reitsma, J. B., Ioannidis, J. P., Macaskill, P., Steyerberg, E. W., ... & Collins, G. S. (2015). Transparent Reporting of a multivariable prediction model for Individual Prognosis or Diagnosis (TRIPOD): explanation and elaboration. Annals of internal medicine, 162(1), W1-W73.